# Estimation of Effects of Sequential Treatments by Reparameterizing Directed Acyclic Graphs

**James M. Robins**
Departments of Epidemiology
and Biostatistics
Harvard School of Public Health
677 Huntington Avenue
Boston, MA 02115

**Larry Wasserman**
Department of Statistics
Carnegie Mellon University
Pittsburgh, PA 15213

## Abstract

The standard way to parameterize the distributions represented by a directed acyclic graph is to insert a parametric family for the conditional distribution of each random variable given its parents. We show that when one's goal is to test for or estimate an effect of a sequentially applied treatment, this natural parameterization has serious deficiencies. By reparameterizing the graph using structural nested models, these deficiencies can be avoided.

## 1 INTRODUCTION

Consider a set of random variables $V = (X_1, \ldots, X_M)$ whose joint density $f(v)$ is represented by a Directed Acyclic Graph (DAG) $G$. If $Pa_m$ represents the parents of $X_m$, then the density factorizes as

$$f(v) = \prod_{m=1}^{M} f(x_m | pa_m). \qquad (1)$$

In practice, in order to estimate $f(v)$ from independent realizations $V_i, i = 1, \ldots, n$, obtained on $n$ study subjects, one often needs to assume some particular parametric form for each $f(x_m | pa_m)$. Thus one writes $f(v) = \prod_{m=1}^{M} f(x_m | pa_m; \theta_m)$. For example, suppose the parent of $X_2$ is $X_1$. Then $p(x_2 | pa_2; \theta_2)$ might be $N(\beta_0 + \beta_1 x_1, \sigma^2)$ so that $\theta_2 = (\beta_0, \beta_1, \sigma)$. In general, if one inserts a parametric family into the right hand side of each term of (1) and the $\theta_m$ are variation-independent, we call this a standard parameterization of the DAG. This seems to be the usual way of using DAGS in practice. The parameters $\theta_m$ are variation-independent if parameter space for $\theta = (\theta'_1, \ldots, \theta'_M)'$ is the product space $\Theta_1 \times \Theta_2 \ldots \times \Theta_M$ with $\Theta_j$ the parameter space for $\theta_j$.

As natural as it seems to parameterize a DAG in this way, there are problems with the standard parameterization when one's goal is to test for or estimate an effect of a treatment or control variable administered sequentially over time. This has been noted by Robins (1989, 1997a) who proposed "structural nested models" (SNM) to avoid these problems. The next section gives a simple example which illustrates the problem.

Briefly put, the problem is this: Suppose the DAG $G$ represents treatments and covariates in a longitudinal study. Further suppose that the partial ordering of the variables in $V$ entailed by the DAG $G$ is consistent with the temporal ordering of the variables. Under certain conditions, the null hypothesis of "no treatment effect," although identifiable based on the observed data, cannot be tested simply by testing for the presence or absence of arrows in the DAG $G$ as one might expect. These conditions, far from being pathological, are indeed likely to hold in most real examples. Fortunately, the null hypothesis can be tested by examining a particular integral called the "G-functional". The null is true if this integral satisfies a certain complex condition. However, we prove in Theorem 2 that there is an additional complication. Specifically, common choices for the parametric families in a standard parameterization often lead to joint densities such that the integral can never satisfy the required condition; as a consequence, in large samples, the null hypothesis of no treatment effect, even when true, will be falsely rejected regardless of the data. These problems are exacerbated in high dimensional problems where SNMs appear to be the only practical approach. This paper focuses on frequentist methods but the same issues arise if Bayesian methods are used.

### 1.1 An Example

To illustrate the problem we are concerned with, consider the following generic example of a sequential randomized clinical trial depicted by DAG 1a in which data have been collected on variables $(A_0, A_1, L, Y)$ on each of $n$ study subjects. The continuous variables $A_0$ and $A_1$ represent the dose in milligrams of AZT treatment received by AIDS patients at two different times, $t_0$ and $t_1$; the dichotomous variable $L$ records whether a patient was anemic just prior to $t_1$; the continuous variable $Y$ represents a subject's

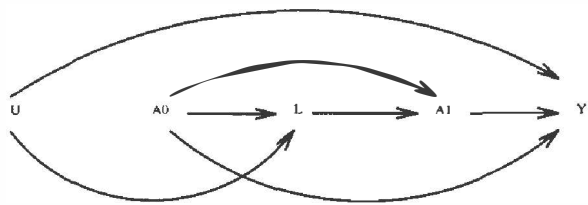

DAG 1a.

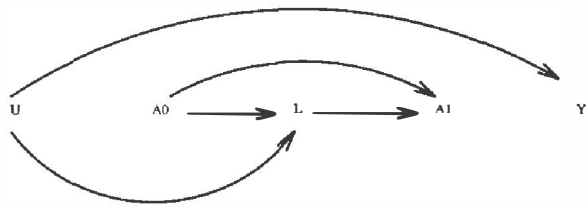

DAG 1b.

HIV-viral load measured at end-of-follow-up; and the hidden (unmeasured) variable $U$ denotes a patient's underlying immune function at the beginning of the study. $U$ is therefore a measure of a patient's underlying health status.

The dose $A_0$ was assigned at random to subjects at time $t_0$ so, by design, $A_0 \coprod U$. Treatment $A_1$ was randomly assigned at time $t_1$ with randomization probabilities that depend on the observed past $(A_0, L)$, so, by design $U \coprod A_1 \mid A_0, L$. For simplicity, we shall assume that no other unmeasured common causes (confounders) exist. That is, each arrow in DAG 1a represents the direct causal effect of a parent on its child, as in Pearl and Verma (1991) or SGS (1993). Note DAG 1a is not complete because of three missing arrows: the arrows from $U$ to $A_0$ and $A_1$ and the arrow from $L$ to $Y$. The arrows from $U$ are missing by design. The missing arrow from $L$ to $Y$ represents a priori biological knowledge that $L$ has no effect on HIV viral load $Y$. (The missing arrow from $L$ to $Y$ is not essential to what follows and is assumed to simplify exposition.) Hence, by the Markov properties of a DAG, we know that $L \coprod Y \mid A_0, A_1, U$. It is also known that AZT $A_0$ causes anemia, so $A_0 \not\coprod L$. Also it is known that the unmeasured variable $U$ has a direct effect on $L$ and $Y$. For example, $U$ causes both anemia and an elevated HIV RNA.

### 1.2 Representing the Null Hypothesis

Suppose the trial data have been collected in order to test the null hypothesis that AZT treatment $(A_0, A_1)$ has no effect on viral load $Y$. The "no AZT effect null" hypothesis is the hypothesis that both the arrow from $A_0$ to $Y$ and the arrow from $A_1$ to $Y$ in DAG 1a are missing, which would imply that the true causal graph generating the data was DAG 1b.

The alternative to this null hypothesis is that the true causal graph generating the data is one of the three

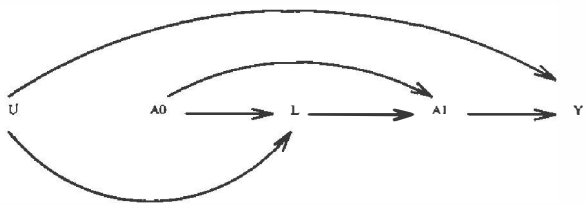

DAG 1c.

graphs 1a, 1c, 1d. Following Pearl and Verma (1991) and SGS (1993), we assume the joint distribution of $W = (U, A_0, A_1, L, Y)$ is faithful to the true graph. That is, if $B, C$, and $D$ are distinct (possibly empty) subsets of the variables in $W$, then $B$ is independent of $C$ given $D$ if and only if $B$ is d-separated from $C$ given $D$ on the true causal graph generating the data. It follows that the "no AZT null hypothesis" of DAG 1b is true if and only if $(A_0, A_1) \coprod Y \mid L, U$. Indeed, since we have assumed no arrows from $L$ to $Y$, $(A_0, A_1) \coprod Y \mid L, U$ is equivalent to the hypothesis $(A_0, A_1) \coprod Y \mid U$. The question is: can we still characterize the null hypothesis even if $U$ is not observed. The answer is yes, according to the following Theorem proved in Sec. 2 below. It also follows from earlier results of Robins (1986) and Pearl and Robins (1995).

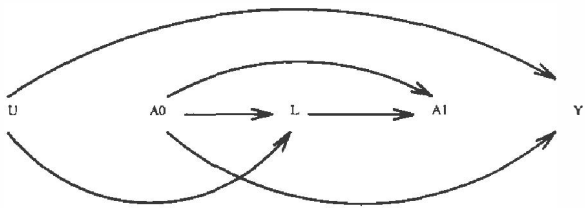

DAG 1d.

**Theorem 1** *Suppose the distribution of $W$ is faithful to DAGs 1a, 1b, 1c, or 1d. Then, the null hypothesis $(A_0, A_1) \coprod Y \mid L, U$ holds if and only if*

$$Y \coprod A_1 \mid A_0, L, \quad i.e. \ f(y \mid \ell, a_0, a_1) = f(y \mid \ell, a_0) \quad (2)$$

*and*

$$\sum_{\ell=0}^{1} f(y \mid a_0, \ell) f(\ell \mid a_0) \text{ does not depend on } a_0. \quad (3)$$

Thus, even though $U$ is unobserved, we can still tell if the null holds by checking (2) and (3) which only involve the observables. Even without imposing faithfulness, $(A_0, A_1) \coprod Y \mid L, U$ implies (2)-(3), although the converse is no longer true.

Consider now the marginal distribution of the observed data $V = (A_0, A_1, L, Y)$. By the d-separation criterion applied to DAG 1a and 1c, we see that if either of DAGs 1a or 1c generated the data, then the joint distribution of $V$ is represented by the complete DAG 2a without missing arrows. If, on the other hand, either

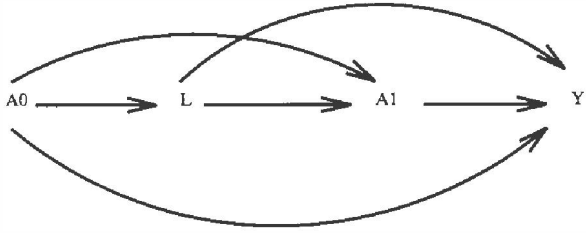

DAG 2a.

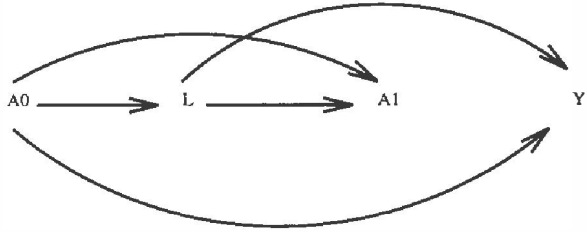

DAG 2b.

DAG 1b or 1d generated the data, then Eq. (2) is true, and the joint distribution of $V$ is represented by the DAG 2b with no arrow from $A_1$ to $Y$. The additional restriction (3) that distinguishes the no effect AZT null hypothesis of graph 1b from DAG 1d is not representable by removing further arrows from DAG 2a. This is an important observation because the most common way of testing whether $(A_0, A_1)$ affects $Y$ is to test for the absence of arrows from $A_0$ to $Y$ and from $A_1$ to $Y$, i.e., to test $(A_0, A_1) \coprod Y \mid L$; we call this the "naive test." This test is incorrect. Specifically, if the no AZT effect hypothesis of DAG 1b is correct and the distribution of $W$ is faithful, then $(A_0, A_1) \coprod Y \mid L$ will be false, and the naive test will falsely reject the no effect AZT null with probability converging to one in large samples.

Thus, testing the null hypothesis of no AZT treatment effect cannot be accomplished by testing for the presence or absence of arrows on DAG 2a. This is because the arrows on the marginal DAG 2a do not have a causal interpretation (even though the arrows on the underlying causal DAG do have a causal interpretation). One solution to this problem is to test (2) and (3) directly. With standard parameterizations, this approach will also fail, as the next section shows.

### 1.3 The Problem With Standard Parameterizations

We saw that to test the null hypothesis, it does not suffice to test whether arrows in DAG 2a from $A_0$ to $Y$ and $A_1$ to $Y$ are broken. Rather, we need to test the conditions (2) and (3). We now show that this test will falsely reject if one uses a standard parameterization. To test the joint null hypothesis (2) and (3), the standard approach is to first specify parametric models for the conditional distribution of each parent given its children in the complete DAG 2a representing the observed data. Hence let $\{f(y \mid a_0, a_1, \ell; \theta); \theta \in \Theta \subset R^q\}$ and $\{f(\ell \mid a_0; \gamma); \gamma \in \Gamma \subset R^p\}$ denote parametric models for the unknown densities $f(y \mid a_0, a_1, \ell)$ and $f(\ell \mid a_0)$. Of course, we cannot guarantee these models are correctly specified. We say the model $f(y \mid a_0, a_1, \ell; \theta)$ is correctly specified if there exists $\theta_0 \in \Theta$ such that $f(y \mid a_0, a_1, \ell; \theta_0)$ is equal to the true (but unknown) density $f(y \mid a_0, a_1, \ell)$ generating the data. Results in this Section require the concept of linear faithfulness. We say that the distribution of $W$ is linearly faithful to the true causal graph generating the data, if for any disjoint (possibly empty) subsets $B$, $C$, and $D$ of the variables in $W$, $B$ is d-separated from $C$ given $D$ on the graph if and only if the partial correlation matrix $r_{BC.D}$ between $B$ and $C$ given $D$ is the zero matrix. If $W$ is jointly normal, linear faithfulness and faithfulness are equivalent. For $W$ non-normal, neither implies the other. However, the argument that the distribution of $V$ should be linearly faithful to the generating causal DAG is essentially identical to the argument that the distribution should be faithful to the causal DAG given by SGS (1993) and Pearl and Verma (1991).

To see why standard parameterizations may not work, consider a specific example. Recall that $Y$ is continuous and that $L$ is binary. Commonly used models in these cases are normal linear regression models and logistic regression models. Thus suppose that we adopt the following models:

$$Y \mid a_0, a_1, \ell; \theta, \sigma^2 \sim N(\theta_0 + \theta_1 a_0 + \theta_2 \ell + \theta_3 a_1, \sigma^2) \quad (4)$$

and

$$f(\ell = 1 \mid a_0; \gamma) = expit(\gamma_0 + \gamma_1 a_0) \quad (5)$$

where $expit(b) = e^b/(1 + e^b)$ and $N(\mu, \sigma^2)$ denotes a Normal distribution with mean $\mu$ and variance $\sigma^2$

We will now prove the following startling result.

**Lemma 1:** If the no AZT effect null hypothesis represented by DAG 1b is true and the distribution of $W$ is either faithful or linearly faithful to DAG 1b, then model (4) and/or model (5) is guaranteed to be misspecified; that is, the set of distributions $\mathcal{F}_{par}$ for $V$ satisfying (4)-(5) is disjoint from the set $\mathcal{F}_{mar}$ of distributions for $V$ that are marginals of distributions for $W$ that are either faithful or linearly faithful to DAG 1b.

Since model (4) and/or (5) are guaranteed to be misspecified under the no AZT effect null hypothesis, one might expect that tests of the null assuming (4)-(5) will perform poorly. This expectation is borne out by the following theorem.

**Theorem 2** *Suppose (i) the data analyst tests the no AZT effect null hypothesis (2)-(3) using the parametric models (4)-(5) fit by the method of maximum likelihood, (ii) the no AZT effect null hypothesis represented by DAG 1b is true, (iii) the distribution of $W$ is linearly faithful to DAG 1b. Then, with probability converging to 1, the no AZT effect null hypothesis (2)-(3) will be falsely rejected.*

Theorem 2 implies that if we use models (4)-(5), then in large samples, we will reject the no AZT effect null hypothesis, even when true, for nearly all data sets (i.e., with probability approaching 1). That is, by specifying models (4)-(5), we will have essentially rejected the no AZT effect null hypothesis, when true, even before seeing the data!

**Proof of Theorem 2 and Lemma 1:** The following Proof of Theorem 2 also proves Lemma 1. Note Eqs. (2)-(3) together imply that $I(a_0, a_1) = \sum_{\ell=0}^{1} E[Y \mid \ell, a_0, a_1] f(\ell \mid a_0)$ does not depend on $(a_0, a_1)$. Now, under model (4)-(5), the maximum likelihood estimator of $I(a_0, a_1)$ is $I\left(a_0, a_1; \widehat{\theta}, \widehat{\gamma}\right) = \widehat{\theta}_0 + \widehat{\theta}_1 a_0 + \widehat{\theta}_3 a_1 + \left\{\widehat{\theta}_2 e^{\widehat{\gamma}_0 + \widehat{\gamma}_1 a_0}\right\} / \left\{1 + e^{\widehat{\gamma}_0 + \widehat{\gamma}_1 a_0}\right\}$ where the maximum likelihood estimators $\widehat{\theta}, \widehat{\gamma}$ satisfy the normal and logistic score equations $\sum_{i=1}^{n} \left(Y_i - \widehat{\theta} Z_i\right) Z_i = 0$ and $0 = \sum_{i=1}^{n} \{L_i - expit(\widehat{\gamma}_0 + \widehat{\gamma}_1 A_{1i})\} (1, A_{1i})'$ where $Z_i = (1, A_{1i}, L_i, A_{2i})'$, $\theta' = (\theta_0, \theta_1, \theta_2, \theta_3)$, and $\gamma' = (\gamma_0, \gamma_1)$. Further, the probability limits $\theta^*$ and $\gamma^*$ of $\widehat{\theta}$ and $\widehat{\gamma}$ satisfy $E[\{Y_i - \theta^* Z_i\} Z_i] = 0$ and $E\left[\{L_i - expit(\gamma_0^* + \gamma_1^* A_{1i})\}(1, A_{1i})'\right] = 0$, where the expectations are with respect to the true distribution generating the data regardless of whether models (4)-(5) are correctly specified. The MLE $I\left(a_0, a_1; \widehat{\theta}, \widehat{\gamma}\right)$ converges in probability to $I(a_0, a_1; \theta^*, \gamma^*)$. It follows that an analyst using models (4)-(5) fit by maximum likelihood will reject (2)-(3) with probability approaching 1 as $n \to \infty$ if $I(a_0, a_1; \theta^*, \gamma^*)$ depends on $a_0, a_1$. We now prove such a dependence by contradiction.

It is clear that $I(a_0, a_1; \theta^*, \gamma^*)$ does not depend on $(a_0, a_1)$ if and only if either *(i)* $\theta_1^* = \theta_2^* = \theta_3^* = 0$, or *(ii)* $\theta_1^* = \theta_3^* = \gamma_1^* = 0$. However, it follows from standard least squares theory that *(i)* is true if and only if $cov[Y, A_0] = cov[Y, A_1] = cov[Y, L] = 0$. But, for example, $cov[Y, L] = 0$ contradicts the assumption that the distribution of $W$ is linearly faithful to DAG 1b since $Y$ and $L$ are not d-separated. Similarly, if *(ii)* is true, then $\gamma_1^* = 0$. But an easy calculation shows that $\gamma_1^* = 0$ if and only if $cov(L, A_0) = 0$. However, $cov(L, A_0) = 0$ contradicts the linear faithfulness assumption since $L$ and $A_0$ are not d-separated on DAG 1b. The argument in this last paragraph also proves Lemma 1.

**Remark:** One might conjecture that the problem could be solved by adding a small number of interaction terms to the model. However, using reasoning like that above, one can show that this is not the case.

## 2   THE G-NULL TEST

A better approach to testing the null is based on the following theorem due to Robins (1986).

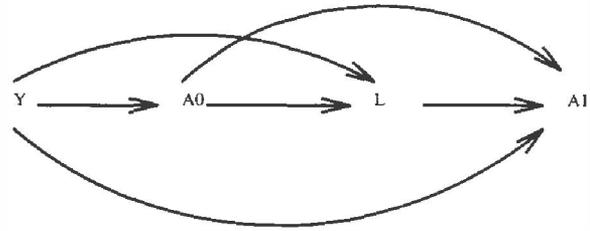

DAG 3a.

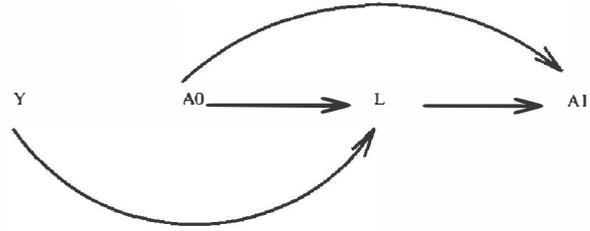

DAG 3b.

**G-Null Theorem:** *Equations (2) and (3) are both true if and only if both (2) and*

$$A_0 \coprod Y \qquad (6)$$

*are true.*

**Remark:** Theorem 1 follows as a corollary since (2) and (6) are all the conditional independences for $V$ entailed by the d-separation rules applied to graph 1b.

From this theorem we see that, under the null (2)-(3), $A_0$ and $Y$ are independent even though there is an arrow from $A_0$ to $Y$ in the marginal DAG 2b for the observed data $V$. In the language of SGS (1993), the distribution is unfaithful to DAG 2b. However, the underlying distribution is not unfaithful to the causally sufficient graph DAG 1b. This is merely a manifestation of the fact that faithfulness need not be preserved under marginalization. SGS's (1993) philosophical argument for faithfulness applies only to the underlying causally sufficient graph in which each arrow has a causal interpretation. It does not apply to marginal subgraphs.

The G-Null Theorem immediately suggests, to those familiar with graphical models, to represent the joint distribution of the observed data by the complete DAG 3a in which the outcome $Y$ comes first followed by $A_0$, then $L$, and finally $A_1$. Then the joint null hypothesis (2) and (3) is represented by DAG 3b in which the arrows from $Y$ to $A_0$ and $Y$ to $A_1$ are removed from the complete DAG 3a. The arrows in DAGs 3a and 3b do not have direct causal interpretations, since, for example, $Y$ is a parent of $L$ even though $L$ is temporally prior to $Y$. Nonetheless, now distributions for $V$ satisfying the no AZT effect null hypothesis (2)-(3) are faithful to the reordered graph 3b.

The "reordering" of DAG 3a is particularly useful in the context of true sequential randomized experiments since then $f(a_0)$ and $f(a_1 \mid a_0, \ell)$ are under the control of the investigator, and thus are known. For example, suppose, by design, $A_0 \sim N(\pi_1, 1)$ and $A_1 \mid A_0, L \sim N(\pi_2(A_0, L), 1)$. Then the models

$$A_0 \mid Y \sim N(\pi_1 + \rho Y, 1)$$

and

$$A_1 \mid A_0, L, Y \sim N(\pi_2(A_0, L) + \rho Y, 1)$$

are known to be correctly specified with the true value of $\rho$ equal to zero under the joint null (2) and (6). Robins (1986, 1992), generalizing Rosenbaum (1984), proposed a g-null test based on the score statistic

$$\frac{\partial}{\partial \rho} \left[ \prod_{i=1}^{n} \log \{ f(A_{2i} \mid A_{1i}, L_i, Y_i; \rho) f(A_{1i} \mid Y_i; \rho) \} \right]_{\mid \rho = 0} \quad (7)$$

which is a sum of bounded independent and identically distributed random variables $U_i = Y_i \{A_{2i} - \pi_2(A_{1i}, L_i)\} + Y_i \{A_{1i} - \pi_1\}$ that have mean zero under the joint null (2) and (6). Therefore, $\chi \equiv \sum_i U_i / \{\sum_i U_i^2\}^{\frac{1}{2}}$ is asymptotically distributed $N(0, 1)$ under the joint null, i.e., under the hypothesis that the distribution of $V$ is represented by DAG 3b. Thus the test that rejects when $|\chi| > 1.96$ is an asymptotically .05-level test of the joint null hypothesis (2) and (6) whatever be the unrestricted, unknown components $f(\ell \mid a_0, y)$ and $f(y)$ of the joint distribution of observables $(A_0, A_1, L, Y)$.

We now have a valid test for the no AZT effect null, but ultimately we want more. In particular, we would like to estimate the size of the treatment effect. To discuss this, we first need to generalize the simple example and then precisely define the treatment effect. We do this in sections 3 - 5. Then we introduce structural nested models which provide a unified approach to estimation of and testing for an AZT treatment effect while avoiding the problems of standardly parameterized DAGs.

## 3 The G-computation Algorithm Formula

Let $G$ be a directed acyclic graph with a vertex set of random variables $V = (V_1, \ldots, V_M)$ with associated distribution function $F(v)$ and density function $f(v)$ with respect to the dominating measure $\mu$. Here $\mu$ is the product measure of Lebesgue and counting measure corresponding to the continuous and discrete components of $V$. By the defining Markov property of DAGs, the density of $V$ can be factored $\prod_{j=1}^{M} f(v_j \mid pa_j)$ where $pa_j$ are realizations of parents $Pa_j$ of $V_j$ on $G$.

We assume $V$ is partitioned into disjoint sets $A$ and $L$ where $A$ equals $\{A_0, \ldots, A_K\}$ are temporally ordered treatments or control variables and given at times $t_0, \ldots, t_K$. $L = \{L_0, L_1, \ldots, L_{K+1}\}$ are response variables. The response variables $L_m$ are temporally subsequent to $A_{m-1}$ and prior to $A_m$. Now for any variable $Z$, let $\mathcal{Z}$ be the support (i.e., the possible realizations) of $Z$. For any $z_0, \ldots, z_m$, define $\overline{z}_m = (z_0, \ldots, z_m)$. By convention $\overline{z}_{-1} \equiv z_{-1} \equiv 0$. Now define a treatment regime or plan $g$ to be a collection of $K + 1$ functions $g = \{g_0, \ldots, g_K\}$ where $g_m : \overline{\mathcal{L}}_m \to \mathcal{A}_m$ maps outcome histories $\overline{\ell}_m \in \overline{\mathcal{L}}_m$ into a treatment $g_m(\overline{\ell}_m) \in \mathcal{A}_m$. If $g_m(\overline{\ell}_m)$ is a constant, say $a_m^*$, not depending on $\overline{\ell}_m$ for each $m$, we say regime $g$ is non-dynamic and write $g = \overline{a}^*$, $\overline{a}^* \equiv (a_0^*, \ldots, a_K^*)$. Otherwise, $g$ is dynamic. We let $\mathcal{G}$ be the set of all regimes $g$

Associated with each regime $g$ is the "manipulated" graph $G_g$ and distribution $F_g(v)$ with density $f_g(v)$ (SGS, 1993). Given the regime $g = (g_0, g_1, \ldots, g_K)$ and the joint density

$$f(v) = f(\ell_0) f(a_0 \mid \ell_0) \cdots f(\ell_{K+1} \mid \overline{\ell}_K, \overline{a}_K), \quad (8)$$

$f_g(v)$ is the density $f(v)$ except that in the factorization (8), $f(a_0 \mid \ell_0)$ is replaced by a degenerate distribution at $a_0 = g_0(\ell_0)$, $f(a_1 \mid \ell_1, a_0, \ell_0)$ is replaced by a degenerate distribution at $a_1 = g_1(\ell_0, \ell_1)$, and, in general, $f(a_k \mid \overline{\ell}_k, \overline{a}_{k-1})$ is replaced by a degenerate distribution at $a_k = g_k(\overline{\ell}_k)$.

Henceforth we shall assume the outcome of interest is $L_{K+1}$ which is assumed to be univariate and shall be denoted by $Y$. In the following, let $g(\overline{\ell}_k) \equiv (g_0(\overline{\ell}_0), \ldots, g_k(\overline{\ell}_k))$ and $g_k(\overline{\ell}_k)$ denote realizations of $\overline{A}_k$ and $A_k$ respectively. Then the marginal density $f_g(y)$ of $Y$ under the distribution $F_g(\cdot)$ is

$$\begin{aligned} f_g(y) &= \int f_g(y, \overline{\ell}_K) d\mu(\overline{\ell}_K) \\ &\equiv \int \{ f_g(y \mid \overline{\ell}_K, g(\overline{\ell}_K)) \qquad (9) \\ &\quad \times \prod_{j=0}^{K} f(\ell_j \mid \overline{\ell}_{j-1}, g(\overline{\ell}_{j-1})) \} d\mu(\ell_j). \end{aligned}$$

Similarly, the marginal distribution function of $Y$ under $F_g(\cdot)$ is

$$F_g(y) = \int \cdots \int pr[Y < y \mid \overline{\ell}_K, g(\overline{\ell}_K)] \\ \times \prod_{j=0}^{K} f(\ell_j \mid \overline{\ell}_{j-1}, g(\overline{\ell}_{j-1})) d\mu(\ell_j). \quad (10)$$

Robins (1986) referred to (10) as the G-computation algorithm formula or functional for the effect of regime $g$ on outcome $Y$. Robins (1986) and Pearl and Robins (1995) give sufficient conditions under which (10) is the distribution of $Y$ that would be observed if all subjects were treated with (i.e., forced to follow) plan $g$. A sufficient condition is that, as in DAG 1a, any hidden variable $U$ that is an ancestor of $A_k$ on the causally sufficient graph generating the data is, for each $k$, d-separated from $A_k$ conditional on the past $(\overline{L}_k, \overline{A}_{k-1})$. This d-separation criteria will be met in any sequential randomized trial and is assumed to hold throughout the remainder of the paper.

## 4 The "g"-null hypothesis

In many settings, the treatments $\overline{A} \equiv \overline{A}_K = (A_0, \ldots, A_K)$ represent a single type of treatment given at different times. In that case, with $Y$ the outcome of interest, an important first question is whether the "g"-null hypothesis of no effect of treatment on $Y$ is true, i.e., whether

$$F_{g_1}(y) = F_{g_2}(y) \text{ for all } y, \text{ and all } g_1, g_2 \in \mathcal{G}. \quad (11)$$

If (11) is true, then the distribution of $Y$ will be the same under any choice of regime $g$, and thus it does not matter whether the treatments $A_k$ are given or withheld at each occasion $k$. One might be concerned that even if (11) is true, the apparently stronger hypothesis that

$$F_{g_1}\left(y \mid \overline{\ell}_k\right) = F_{g_2}\left(y \mid \overline{\ell}_k\right) \quad (12)$$

for all $\overline{\ell}_k$, $y$, and $g_1, g_2 \in \mathcal{G}$ might be false, and so, conditional on $\overline{\ell}_k$, it might matter which regime is to be followed subsequently. However, it is easy to show that the "g"-null hypothesis (11) is equivalent to (12). Here

$$F_g\left(y \mid \overline{\ell}_m\right) = \int pr\left[Y < y \mid \overline{\ell}_K, g\left(\overline{\ell}_K\right)\right]$$
$$\times \prod_{j=m+1}^{K} f(\ell_j \mid \overline{\ell}_{j-1}, g(\overline{\ell}_{j-1})) d\mu(\ell_j). \quad (13)$$

Nevertheless, the "g"-null hypothesis is not implied by the weaker condition that $F_{g=(\overline{a}_1)}(y) = F_{g=(\overline{a}_2)}(y)$ for all non-dynamic regimes $\overline{a}_1 \equiv \overline{a}_{1K}$ and $\overline{a}_2$. However, the following lemma is true. The Lemma restates the "g"-null hypothesis in terms of restrictions on the conditional distributions $F_g\left(y \mid \overline{\ell}_k\right)$ for non-dynamic regimes $g$.

**Lemma:** *The "g"-null hypothesis is true if and only if $F_{g=(\overline{a}_1)}\left(y \mid \overline{\ell}_k\right) = F_{g=(\overline{a}_2)}\left(y \mid \overline{\ell}_k\right)$ for all $y$, $\overline{\ell}_k$, $\overline{a}_1$ and $\overline{a}_2$, with $\overline{a}_1$ and $\overline{a}_2$ agreeing through occasion $t_{k-1}$, i.e., $\overline{a}_{1(k-1)} = \overline{a}_{2(k-1)}$.*

If we apply this Lemma to the simple example in Section 1, we recover (2) and (3). That is, the "g"-null hypothesis for the observed data $V = (A_0, A_1, L, Y)$ of Sec. 1 is precisely (2)-(3).

### 4.1 Failure of the usual parameterization for testing the "g"-null hypothesis

In Section 1, we saw that is was difficult to test the "g"-null hypothesis using the usual parameterization of a DAG. These problems are exacerbated in the general case. Indeed, there are several difficulties. First, even if the densities appearing in the G-computation formula (10) were known for each $g \in \mathcal{G}$, since $F_g(y)$ is a high-dimensional integral, in general, it cannot be analytically evaluated for any $g$ and thus, must be evaluated by a Monte Carlo integral approximation — the Monte Carlo G-computation algorithm (Robins, 1987, 1989). Second, even if $F_g(y)$ could be well-approximated for each regime $g$, the cardinality of the set $\mathcal{G}$ is enormous [growing at faster than an exponential rate in $K$ (Robins, 1989)]. Thus it would be computationally infeasible to evaluate $F_g(y)$ for all $g \in \mathcal{G}$ to determine whether the "g"-null hypothesis held.

However, as we saw in Sec. 1, the most fundamental difficulty with the usual parameterization of a DAG in terms of the densities $f(v_j \mid pa_j)$ is that it is only sufficient but not necessary for the "g"-null hypothesis to hold that $f(\ell_j \mid \overline{\ell}_{j-1}, \overline{a}_{j-1})$ and $f(y \mid \overline{\ell}_K, \overline{a}_K)$ do not depend on $\overline{a}_{j-1}$ and $\overline{a}_K$ respectively. As a consequence, if we use standard parametric models for $f(v_j \mid pa_j)$, *(i)*there is no parameter, say $\psi$, which takes the value zero if and only if the "g"-null hypothesis is true, and *(ii)* the "g"-null hypothesis, even when true, may, with probability approaching 1, be rejected in large samples.

## 5 G-null Tests

As in the special case discussed in Sec. 1, a better approach to testing the "g"-null hypothesis is based on the following theorem of Robins (1986).

**G-null theorem:** The "g"-Null Hypothesis (11) is true $\Leftrightarrow$

$$Y \coprod A_k \mid \overline{L}_k, \overline{A}_{k-1}, \quad k = (0, \ldots, K). \quad (14)$$

We now use (14) to construct g-null tests. For variety, in this section we shall suppose $A_k$ is dichotomous. Suppose we can correctly specify a logistic model

$$f\left(A_m = 1 \mid \overline{L}_m, \overline{A}_{m-1}\right) = \{1 + \exp\left(-\alpha'_0 W_m\right)\}^{-1}, \quad (15)$$

$m = 0, \ldots, K$, where $W_m$ is a known $p$-dimensional function of $\overline{L}_m, \overline{A}_{m-1}$. This will always be possible in a true sequential randomized trial since $f\left(A_m = 1 \mid \overline{L}_m, \overline{A}_{m-1}\right)$ is known by design.

Let $Q_m \equiv q\left(Y, \overline{L}_m, \overline{A}_{m-1}\right)$ where $q(\cdot, \cdot, \cdot)$ is any known real-valued function chosen by the data analyst. Let $\theta$ be the coefficient of $Q_m$ when $\theta Q_m$ is added to the regressors $\alpha'_0 W_m$ in (15). If, for each $m$, (15) is true for some $\alpha_0$, then hypothesis (14) is equivalent to the hypothesis the true value $\theta_0$ of $\theta$ is zero. A score, Wald, or likelihood ratio test of the hypothesis $\theta_0 = 0$ can then be computed using logistic regression software where, when fitting the logistic regression model, each subject is regarded as contributing $K + 1$ independent Bernoulli observations $A_{m,i}$ - one at each treatment time $t_0, t_1, \ldots, t_K$. Robins (1992) refers to any such test as a g-test and provides mathematical justification. A g-test is a semiparametric test since it only requires we specify a parametric model for $f\left(A_m \mid \overline{L}_m, \overline{A}_{m-1}\right)$ rather than for the entire joint distribution of the observed data $V = \left(\overline{L}_{K+1}, \overline{A}_K\right)$. In a true sequential randomized trial $\alpha_0$ will be known and need not be estimated. In an observational study, $\alpha_0$ will need to be estimated, and the g-test is only guaranteed to reject at its nominal level if the model (15) is correct. However, in contrast to the disturbing results summarized in Lemma 1, any parametric model $f\left(a_m \mid \bar{\ell}_m, \bar{a}_{m-1}; \alpha\right)$ is compatible with the "g"-null hypothesis (11). That is, there exist joint distributions for $V$ under which the parametric model $f\left(a_m \mid \bar{\ell}_m, \bar{a}_{m-1}; \alpha\right)$ is correctly specified and the "g"-null hypothesis (11) holds.

## 6  Structural Nested Models

In this Section, we describe the class of structural nested models. In this paper, we shall only consider the simplest structural nested model - a structural nested distribution model for a univariate continuous outcome $Y$ measured after the final treatment time $t_K$. Robins (1989, 1992, 1994, 1995) considers generalizations to discrete outcomes, multivariate outcomes, and failure time outcomes.

We assume $L_{K+1}$ is a univariate continuous-valued random variable with a continuous distribution function and denote it by $Y$. Our g-test of the "g"-null hypothesis (11) was unlinked to any estimator of $F_g(y)$. Our first goal in this subsection will be to derive a complete reparameterization of the joint distribution of $V$ that will offer a unified fully parametric likelihood-based approach to testing the "g"-null hypothesis and estimating the function $F_g(y)$. Then we will develop a unified approach to testing and estimation based on the semiparametric g-test of Sec. 5.

### 6.1  A New Characterization of the "g"-Null Hypothesis

The first step in constructing our reparameterization of the distribution of $V$ is a new characterization of the "g"-null hypothesis (11). We assume the conditional distribution of $Y$ given $(\bar{\ell}_m, \bar{a}_m)$ has a continuous positive density with respect to Lebesgue measure. Given any treatment history $\bar{a} = \bar{a}_K$, adopt the convention that $(\bar{a}_m, 0)$ is the treatment history that agrees with $\bar{a}$ through $t_m$ and is zero thereafter. Recall that the quantile-quantile function $\gamma(y)$ linking any two continuous distribution functions $F_1(y)$ and $F_2(y)$ is $\gamma(y) = F_1^{-1}\{F_2(y)\}$. It maps quantiles of $F_2(y)$ into quantiles of $F_1(y)$.

Let $\gamma(y, \bar{\ell}_m, \bar{a}_m)$ be the quantile-quantile function mapping quantiles of $F_{g=(\bar{a}_m, 0)}(y \mid \bar{\ell}_m)$ into quantiles of $F_{g=(\bar{a}_{m-1}, 0)}(y \mid \bar{\ell}_m)$.

It follows from its definition as a quantile-quantile function that: (a) $\gamma(y, \bar{\ell}_m, \bar{a}_m) = y$ if $a_m = 0$; (b) $\gamma(y, \bar{\ell}_m, \bar{a}_m)$ is increasing in $y$; and (c) the derivative of $\gamma(y, \bar{\ell}_m, \bar{a}_m)$ w.r.t. $y$ is continuous. Examples of such functions are

$$\gamma(y, \bar{\ell}_m, \bar{a}_m) = y + 2a_m + 3a_m a_{m-1} + 4a_m w_m \quad (16)$$

where $w_m$ is a given *univariate function* of $\bar{\ell}_m$ and

$$\gamma(y, \bar{\ell}_m, \bar{a}_m) = y \exp\{2a_m + 3a_m a_{m-1} + 4a_m w_m\}. \quad (17)$$

Our interest in $\gamma(y, \bar{\ell}_m, \bar{a}_m)$ is based on the following theorem proved in Robins (1989, 1995a).

**Theorem 3** $\gamma(y, \bar{\ell}_m, \bar{a}_m) = y$ *for all* $y, m, \bar{\ell}_m, \bar{a}_m$ *if and only if the "g"-null hypothesis (11) holds.*

### 6.2  Pseudo-Structural and Structural Nested Distribution Models

In view of theorem (3), our approach will be to construct a parametric model for $\gamma(y, \bar{\ell}_m, \bar{a}_m)$ depending on a parameter $\psi$ such that $\gamma(y, \bar{\ell}_m, \bar{a}_m) = y$ if and only if the true value $\psi_0$ of the parameter is 0. We will then reparameterize the density of the observables $V$ in terms of a random variable which is a function of the observables and the function $\gamma(y, \bar{\ell}_m, \bar{a}_m)$. As a consequence, likelihood-based tests of the hypothesis $\psi_0 = 0$ will produce likelihood-based tests of the "g"-null hypothesis.

Definition: The distribution $F$ of $V$ follows a pseudo-structural nested distribution model $\gamma^*(y, \bar{\ell}_m, \bar{a}_m, \psi)$ if $\gamma(y, \bar{\ell}_m, \bar{a}_m) = \gamma^*(y, \bar{\ell}_m, \bar{a}_m, \psi_0)$ where (1) $\gamma^*(\cdot, \cdot, \cdot, \cdot)$ is a known function; (2) $\psi_0$ is a finite vector of unknown parameters to be estimated; (3) for each value of $\psi$, $\gamma^*(y, \bar{\ell}_m, \bar{a}_m, \psi)$ satisfies the conditions *(a)*, *(b)*, and *(c)* that were satisfied by $\gamma(y, \bar{\ell}_m, \bar{a}_m)$; (4) $\partial \gamma^*(y, \bar{\ell}_m, \bar{a}_m, \psi)/\partial \psi'$ and $\partial^2 \gamma^*(y, \bar{\ell}_m, \bar{a}_m, \psi)/\partial \psi' \partial y$ are continuous for all $\psi$; and (5) $\gamma^*(y, \bar{\ell}_m, \bar{a}_m, \psi) = y$ if and only if $\psi = 0$ so that $\psi_0 = 0$ represents the "g-" null hypothesis.

Examples of appropriate functions $\gamma^*(y, \bar{\ell}_m, \bar{a}_m, \psi)$ can be obtained from Eqs. (16) and (17) by replacing the quantities 2, 3 and 4 by the components of $\psi = (\psi_1, \psi_2, \psi_3)$. We call models for $\gamma(y, \bar{\ell}_m, \bar{a}_m)$ pseudo-structural because pseudo-SNDMs are models for the distribution $F$ of the observables $V$ regardless of whether this distribution has a structural (i.e. causal) interpretation (as it would in a sequential randomized trial). When $\gamma(y, \bar{\ell}_m, \bar{a}_m)$ does have a causal interpretation as well, we refer to our models as structural nested distribution models (SNDMs).

### 6.3  The Reparameterization of the Likelihood

Next we recursively define random variables $H_K, \ldots, H_0$ that depend on the observables $V$ as follows. $H_K \equiv \gamma(Y, \overline{L}_K, \overline{A}_K)$, $H_m \equiv h_m(Y, \overline{L}_K, \overline{A}_K) \equiv \gamma(H_{m+1}, \overline{L}_m, \overline{A}_m)$, and $H \equiv h(Y, \overline{L}_K, \overline{A}_K) \equiv H_0$. Note $Y = h^{-1}(H, \overline{L}_K, \overline{A}_K)$ is a deterministic function of $H, \overline{L}_K, \overline{A}_K$ where for any function $q(y, \bullet)$, we define $q^{-1}(u, \bullet) = y$ if $q(y, \bullet) = u$. Note by Theorem (3) if the "g"-null hypothesis is true, then $H = Y$.

Example: If $\gamma(y, \bar{\ell}_m, \bar{a}_m)$ is given by Eq. (16), then
$H_m = Y + \sum_{k=m}^{K}(2A_k + 3A_k A_{k-1} + 4A_k W_k)$ and $h^{-1}(u, \bar{\ell}_K, \bar{a}_K) = u - \left[\sum_{m=0}^{K} 2a_m + 3a_m a_{m-1} + 4a_m w_m\right]$.

Since $\gamma(Y, \overline{L}_m, \overline{A}_m)$ is increasing in $Y$, the map from $V \equiv (Y, \overline{L}_K, \overline{A}_K)$ to $(H, \overline{L}_K, \overline{A}_K)$ is 1-1 with a strictly positive Jacobian determinant. Therefore, $f_{Y,\overline{L}_K,\overline{A}_k}(Y, \overline{L}_K, \overline{A}_K) = \{\partial H/\partial Y\} f_{H,\overline{L}_K,\overline{A}_K}(H, \overline{L}_K, \overline{A}_K)$. However, Robins (1989, 1995) proves that
$$A_m \coprod H \mid \overline{L}_m, \overline{A}_{m-1}. \qquad (18)$$
It follows that
$$f_{Y,\overline{L}_K,\overline{A}_K}(Y, \overline{L}_K, \overline{A}_K) = \{\partial H/\partial Y\} f\{H\}$$
$$\prod_{m=0}^{K} f(L_m \mid \overline{L}_{m-1}, \overline{A}_{m-1}, H) f[A_m \mid \bar{L}_m, \bar{A}_{m-1}]. \qquad (19)$$
(19) is the aforementioned reparameterization of the density of the observables. For completeness, we prove (18) in the Appendix.

**Remark**: Eq. (19) is only a reparameterization. In particular, Eqs. (18) and (19) do not translate into restrictions on the joint distribution of $V = (Y, \overline{L}_K, \overline{A}_K)$ since any law for $V$ satisfies (18) and (19). Conversely, (i) any function $\gamma(y, \bar{\ell}_m, \bar{a}_m)$ satisfying $\gamma(y, \bar{\ell}_m, \bar{a}_m) = y$ if $a_m = 0$, $\partial\gamma(y, \bar{\ell}_m, \bar{a}_m)/\partial y$ is positive and continuous and (ii) any densities $f_H(h)$, $f_{L_m}(\ell_m \mid \bar{\ell}_{m-1}, \bar{a}_{m-1}, h)$ and $f_{A_m}(a_m \mid \bar{\ell}_m, \bar{a}_{m-1})$ together induce a unique law for $V = (Y, \overline{L}_K, \overline{A}_K)$ by (a) using $f_H(\bullet), f_{L_m}(\bullet \mid \bullet)$ and $f_{A_m}(\bullet \mid \bullet)$ to determine a joint distribution for $(H, \overline{L}_K, \overline{A}_K)$ satisfying $H \coprod A_m \mid \overline{A}_{m-1}, \overline{L}_m$ and then (b) defining $Y$ to be $h^{-1}(H, \overline{L}_K, \overline{A}_K)$ with $h^{-1}(\cdot, \cdot, \cdot)$ defined in terms of $\gamma(\cdot, \cdot, \cdot)$ as above.

It follows that joint distribution $F$ for the observed data $V$ can be represented by an enlarged DAG $G_{en\ell}$ based on the ordering $H, L_0, A_0, L_1, A_1, \ldots, L_K, A_K, Y$ which is complete except with arrows from $H$ to $A_k, k = 0, \ldots, K$, missing. The dependence of $Y$ on its parents $(H, \overline{A}_K, \overline{L}_K)$ is completely deterministic. Furthermore, by Theorem 3, the "g"-null hypothesis is represented by the subgraph of this DAG in which all arrows into $Y$ are removed except for the arrow from $H$ [since, under the "g"-null hypothesis, $Y = H$]. Robins (1989, 1995) shows $F_g(y)$ based on $G_{en\ell}$ is equal to $F_g(y)$ based on DAG G if the $g_k(\bullet), k = 0, \ldots, K$, are not functions of $H$.

Thus we have succeeded in reparameterizing the joint density of the observables in terms of the function $\gamma(y, \bar{\ell}_m, \bar{a}_m)$, its derivative with respect to $y$, and the densities $f[\ell_m \mid \bar{\ell}_{m-1}, \bar{a}_{m-1}, h]$, $f(h)$, and $f[a_m \mid \bar{\ell}_m, \bar{a}_{m-1}]$.

We can then specify a fully parametric model for the joint distribution of the observables by specifying (a) a pseudo-SNFTM $\gamma^*(y, \bar{\ell}_m, \bar{a}_m, \psi_0)$ for $\gamma(y, \bar{\ell}_m, \bar{a}_m)$, and (b) parametric models $f[\ell_m \mid \bar{\ell}_{m-1}, \bar{a}_{m-1}, h; \phi_0]$, $f(h; \eta_0)$, and $f[a_m \mid \bar{\ell}_m, \bar{a}_{m-1}; \alpha_0]$ for the above densities. It follows from (19) that the maximum likelihood estimates of $(\phi_0, \eta_0, \psi_0)$ are the values $(\hat{\phi}, \hat{\eta}, \hat{\psi})$ that maximize

$$\prod_{i=1}^{n}\left\{\left[\frac{\partial H_i(\psi)}{\partial Y_i}\right] f(H_i(\psi); \eta)\right\} \\ \times \prod_{m=0}^{m=K} f(L_{m,i} \mid \overline{L}_{m-1,i}, \overline{A}_{m-1,i}, H_i(\psi); \phi) \qquad (20)$$

where $H(\psi)$ is defined like $H$ except $\gamma^*(y, \bar{\ell}_m, \bar{a}_m, \psi)$ replaces $\gamma(y, \bar{\ell}_m, \bar{a}_m)$. Since the "g"-null hypothesis is equivalent to the hypothesis that $\psi_0 = 0$, the reparameterization (19) has allowed us to construct fully parametric likelihood-based tests of the "g"-null hypothesis based on the Wald, score, or likelihood ratio test for $\psi_0 = 0$ based on the likelihood (20).

### 6.4 Estimation of $F_g(y)$

If our fully parametric likelihood-based test of the null hypothesis $\psi_0 = 0$ rejects, we would wish to employ these same parametric models to estimate $F_g(y)$ for each $g \in \mathcal{G}$. We shall accomplish this goal in two steps. First we provide a Monte Carlo algorithm which produces independent realizations $y_{v,g}$ of a random variable whose distribution function is $F_g(y)$.

MC Algorithm: Given a regime $g$:

Step (1): Draw $h_v$ from $f_H(h)$

Step (2): Draw $\ell_{o,v}$ from $f[\ell_o \mid h_v]$

Step (3): Do for $m = 1, \ldots, K$

Step (4): Draw $\ell_{m,v}$ from $f[\ell_m \mid \bar{\ell}_{m-1,v}, g(\bar{\ell}_{m-1,v}), h_v]$.

Step (5): Compute $y_{v,g} = h^{-1}(h_v, \bar{\ell}_{K,v}, g(\bar{\ell}_{K,v}))$.

Robins (1989, 1995) shows that the $(y_{1,g}, \ldots, y_{v,g}, \ldots)$ are independent simulations from $F_g(y)$ based on $G_{en\ell}$ and thus are independent realizations of a random variable with distribution function $F_g(y)$ based on DAG G.

Second, since $f_H(h)$ and $f[\ell_m \mid \bar{\ell}_{m-1,v}, g(\bar{\ell}_{m-1,v}), h_v]$ are unknown, in practice, we replace them with the estimates obtained above.

If $\gamma(y, \bar{\ell}_m, \bar{a}_m) \equiv \gamma(y, \bar{a}_m)$ does not depend on $\bar{\ell}_m$ (i.e., there is no treatment-covariate interaction), then, in the above algorithm for a non-dynamic regime $g = (\bar{a})$, steps (3)-(5) can be eliminated since $h^{-1}(u, \bar{\ell}_K, \bar{a}_K) \equiv h^{-1}(u, \bar{a}_K)$ does not depend on $\bar{\ell}_K$; as a consequence, to draw from $F_{g=(\bar{a})}(y)$ one does not need to model the conditional density of the variables $L_m$. In fact, $F_{g=(\bar{a})}(y) = pr\{h^{-1}[H, \bar{a}_K] > y\}$.

Example: If $\gamma(y, \bar{\ell}_m, \bar{a}_m) = y + 2a_m + 3a_m a_{m-1}$ then

$$h^{-1}(u, \overline{a}_K) = u - \sum_{m=0}^{K} 2a_m + 3a_m a_{m-1}.$$

## 7 Semiparametric Inference in SNMs

### 7.1 g-Estimation of $\psi_0$

In this Section, we assume $A_m$ is dichotomous. Robins (1992) discusses generalizations to multivariate $A_m$ with (possibly) continuous components. Robins (1992) argues that even in observational studies, one will have better prior knowledge about, and thus can more accurately model, the densities $f(A_m = 1 \mid \overline{L}_m, \overline{A}_{m-1})$ [as in Eq. (15)] than the densities occurring in Eq. (20). Indeed, with some loss of efficiency, if $A_k$ and $L_k$ are discrete, we can use a saturated model in Eq. (15), thus eliminating all possibility of misspecification. Additionally, there is no possibility of misspecification in sequential randomized trials since then $f(A_m = 1 \mid \overline{L}_m, \overline{A}_{m-1})$ is under the control of and thus known to the investigator. It is for these reasons we prefer to test the "g"-null hypothesis $\psi_0 = 0$ using the g-test of Sec. 5 rather than the likelihood-based test of Sec. 6.4. Here, we describe how to obtain $n^{\frac{1}{2}}$-consistent g-estimates $\widetilde{\psi}$ of $\psi_0$ which are based on model (15) and thus are consistent with the g-test of $\psi_0 = 0$ in the sense that 95% confidence intervals for $\psi_0$ will fail to cover 0 if and only if the g-test of $\psi_0 = 0$ rejects at the .05 level. Specifically, we (a) add $\theta' Q_m^*(\psi)$ [rather than $\theta Q_m$] to the regressors $\alpha_0' W_m$ in (15) where $Q_m^*(\psi) = q^*\{H(\psi), \overline{L}_m, \overline{A}_{m-1}\}$, $q^*()$ is a known vector-valued function of $\dim \psi$ chosen by the data analyst, $\theta$ is a $\dim \psi$ valued parameter; (b) define the G-estimate $\widetilde{\psi}$ to be the value of $\psi$ for which the logistic regression score test statistic of $\theta = 0$ is precisely zero; and (c) a 95% large sample confidence set for $\psi_0$ is the set of $\psi$ for which the score test (which we call a g-test) of the hypothesis $\theta = 0$ fails to reject (Robins, 1992). The parameter $\psi$ is treated as a fixed constant when calculating the score test. The optimal choice of the function $q^*()$ is given in Robins (1992).

Given $\widetilde{\psi}$, we now estimate $F_g(y)$ by (i) finding $\widetilde{\phi}$ that maximizes (20) with the expression in set braces set to 1 and with $\psi$ fixed at $\widetilde{\psi}$, (ii) using the empirical distribution of $H_i(\widetilde{\psi})$ as an estimate for the distribution of $H$ and (iii) using the MC algorithm of Sec. (6.4) to estimate $F_g(y)$ based on $(\widetilde{\psi}, \widetilde{\phi})$ and the empirical law of $H_i(\widetilde{\psi})$.

Indeed, if $h^{-1}(u, \overline{\ell}_K, \overline{a}_K)$ does not depend on $\overline{\ell}_K$, then a $n^{\frac{1}{2}}$-consistent estimate of $F_{g=(\overline{a})}(y)$ is $n^{-1} \sum_{i=1}^n I\left\{h^{-1}\left[H_i(\widetilde{\psi}), \overline{a}_K, \widetilde{\psi}\right] > y\right\}$ where $I\{A\} = 1$ if $A$ is true and is zero otherwise. This estimator has the distinct computational advantage over an estimator based on the G-computation algorithm formula (10) of requiring neither integration nor modelling of the conditional density of the covariates $L_m$ given the past.

## 8 Testing and estimation of Direct Effects

### 8.1 Inadequacy of SNDMs

Structural nested distribution models are appropriate for testing for and estimation of the joint effect of a single time-dependent treatment $\overline{A}_K$ given sequentially in which the null hypothesis of interest is the "g"-null hypothesis (11). This model is inappropriate for testing the null hypothesis of whether a given treatment (say, $A_0$) has a direct effect on the outcome $Y$ when a subsequent treatment (say, $A_1$) is manipulated (set) to a particular value $a_1$. Appropriate models for direct effects are discussed in Appendix 3 of Robins (1997a) and Robins (1997b). In Section 8.3, we provide an introduction to these models. In this subsection, we demonstrate why SNDM models are not appropriate for testing for direct effects.

Specifically, again consider the example of Sec. 1 but now suppose $A_0$ is the drug aerosolized pentamidine while $A_1$ remains treatment with AZT. Suppose it is known that AZT has a direct effect on the outcome $Y$. Our goal is to test the null hypothesis that there is no direct effect of $A_0$ on $Y$. Our knowledge that $A_1$ affects $Y$ rules out DAGs 1b and 1d as the causal DAGs generating the data. The null hypothesis of no direct effect of $A_0$ on $Y$ is represented by DAG 1c in which the arrow from $A_0$ to $Y$ is missing. The alternative hypothesis is represented by graph 1a. Verma and Pearl (1991) and SGS (1993, p. 192) have also considered this testing problem.

The restriction on the marginal distribution of $V = (A_1, A_2, L, Y)$ entailed by the no $A_0$ effect null hypothesis of DAG 1c is that $f_{g=(a_0, a_1)}(y)$ is not a function of $a_0$ which cannot be represented by a conditional independence constraint amongst subsets of the variables in $V$ (Robins, 1986; Verma and Pearl, 1991; SGS, 1993, p. 193). Robins (1997a) shows this restriction is equivalent to the hypothesis that

$$\gamma(y, a_0, \ell_0) = y \qquad (21)$$

and

$$E\left\{\gamma(y, \overline{a}_1, \overline{L}_1) \mid A_0 = a_0, L_0 = \ell_0\right\} \qquad (22)$$

does not depend on $a_0$. Note that, in our example, $L_0$ is not present and $L_1 \equiv L$. Suppose therefore we choose to test (21) and (22) by specifying (i) a SNDM given by

$$\gamma^*(y, \ell_0, a_0; \psi) \equiv \gamma^*(y, a_0; \psi) = y - \psi_1 a_0 \qquad (23)$$

and

$$\gamma^*(y, \overline{\ell}_1, \overline{a}_1, \psi) \equiv \gamma^*(y, \ell, \overline{a}_1, \psi) \\ = y - \psi_2 a_1 - \psi_3 a_1 a_0 - \psi_4 a_1 \ell - \psi_5 a_1 a_0 \ell \qquad (24)$$

and *(ii)* the logistic model (5) for the probability of $L$ given $A_0$.

To simplify the following argument, we suppose that $U$ is dichotomous. When causal DAG 1c generated the data, we say that there is an $A_1 - U$ treatment interaction if $\gamma^{(1)}(y, a_1) \neq \gamma^{(0)}(y, a_1)$ where $\gamma^{(j)}(y, a_1)$ maps quantiles of $F(y \mid a_1, U = j)$ into those of $F(y \mid a_1 = 0, U = j)$. For example, if $A_1$ affects $Y$ only when $U = 1$ and has no effect when $U = 0$ there is an $A_1 - U$ interaction since then $\gamma^{(0)}(y, A_1) = y$ and $\gamma^{(1)}(y, A_1) \neq y$. If there is an $A_1 - U$ interaction, then, similarly to Lemma 1, either the logistic model (5) and/or the structural nested distribution model (24) must be misspecified under the no $A_0$ effect null hypothesis. Formally, we have

**Lemma 2:** If *(i)* the no $A_0$ effect null hypothesis represented by DAG 1c is true, *(ii)* the distribution of $W$ is either faithful or linearly faithful to DAG 1c, and *(iii)* there is an $A_1 - U$ interaction, then model (5) and/or the SNDM (24) is misspecified.

We conclude that it is not adequate to test for and/or estimate direct effects using either the standard DAG parameterization or the reparameterization induced by a SNDM. Robins (1997a, App. 3) and Robins (1997b) suggest "direct effect" structural nested models which lead to alternative appropriate reparameterizations.

**Proof of Lemma 2:** We noted above that DAG 1c implies (22). Now under our models (5) and (24), (22) can be written

$$y - \psi_2 a_1 - \psi_3 a_0 a_1 - [\psi_4 a_1 + \psi_5 a_0 a_1] \, expit \, [\gamma_0 + \gamma_1 a_0]$$

which does not depend on $a_0$ if and only if either

$$\psi_3 = \psi_5 = \gamma_1 = 0 \qquad (25)$$

or

$$\psi_3 = \psi_4 = \psi_5 = 0. \qquad (26)$$

Now, since $L$ and $A_0$ are not d-separated on DAG 1c, we conclude $\gamma_1 \neq 0$ if the distribution of $W$ is faithful or linearly faithful to DAG 1c.

We now complete the proof by showing that (26) also cannot be true. Were (26) true, we conclude from model (24) that $\gamma(y, \bar{\ell}_1, \bar{a}_1) = \gamma(y, a_1)$ does not depend on $a_0$ or $\ell = \ell_1$, resulting in the following contradiction. Since $(L, A_0) \coprod Y \mid A_1, U$ and $A_1 \coprod U \mid L, A_0$ on DAG 1c, $F(y \mid a_1, a_0, \ell) = F(y \mid a_1, U = 1) p(a_0, \ell) + F(y \mid a_1, U = 0) \{1 - p(a_0, \ell)\}$ where $p(a_0, \ell) = pr[U = 1 \mid a_0, \ell]$. Now by definition of $\gamma(\bar{\ell}_1, \bar{a}_1)$, $F[\gamma(\bar{\ell}_1, \bar{a}_1) \mid a_1 = 0, a_0, \ell] = F[y \mid a_1, a_0, \ell]$. So, if (23) is true, we have $F[\gamma(y, a_1) \mid a_1 = 0, U = 1] p(a_0, \ell) + F[\gamma(y, a_1) \mid a_1 = 0, U = 0][1 - p(a_0, \ell)] = F[y \mid a_1, U = 1] p(a_0, \ell) + F(y \mid a_1, U = 0)[1 - p(a_0, \ell)]$. This implies that $F(y \mid a_1, U = j) = F[\gamma(y, a_1) \mid a_1 = 0, U = j]$ for $j = 0, 1$ which can be rewritten as $\gamma^{(0)}(y, a_1) = \gamma^{(1)}(y, a_1) = \gamma(y, a_1)$ contradicting premise *(iii)*.

### 8.2 Direct-effect g-null test

An appropriate approach to testing the no direct $A_0$ effect null hypothesis is based on the following theorem.

**Theorem 4: Direct-effect G-null theorem:** The no direct $A_0$ effect null hypothesis that $f_{g=(a_0, a_1)}(y)$ does not depend on $a_0$ is true if and only if, for any functions $t_1(\bullet)$ and $t_2(\bullet)$, $E[t_1(Y) t_2(A_1) / W_1 \mid A_0]$ does not depend on $A_0$ w.p.1, whenever the expectation is finite, where $W_1 \equiv f(A_1 \mid L, A_0)$.

**Proof of Theorem 4:** By Fubini's theorem, the expectation can be written $\int_{-\infty}^{\infty} t_2(a_1) q(a_1, A_0) da_1$,

$$q(a_1, A_0) \equiv \int_{-\infty}^{\infty} t_1(y)$$
$$\times \left[ \sum_{\ell=0}^{1} f(y \mid \ell, a_1, A_0) f(\ell \mid A_0) \right] dy. \qquad (27)$$

Now the term in square brackets in (27) is $f_{g=(A_0, a_1)}(y)$. Recalling that $t_1(y)$ is arbitrary proves the theorem.

**Remark 1:** If $A_1$ is discrete, we can always choose $t_2(a_1) \equiv 1$. However, as in this example, if $A_1$ is continuous, we need to choose $t_2(a_1)$ so as to make (27) finite. For example, if $q(a_1, A_0)$ were identically 1, then (27) is finite if and only if $\int_{-\infty}^{\infty} t_2(a_1) da_1$ is finite.

**Remark 2:** If on DAG 1c, $A_1$ had been parentless, then, by d-separation and faithfulness, the no direct $A_0$ effect null hypothesis of no arrow from $A_0$ to $Y$ is true if and only if $A_0$ and $Y$ are independent. Theorem 4 implies that any test of independence of $A_0$ and $Y$ (which is linear in $Y$) can still be used to test the no direct $A_0$ effect null hypothesis when $A_1$ has parents $(A_0, L)$ provided, in calculating the test, $Y$ is replaced by $\mathcal{W} = t_1(Y) t_2(A_1) / W_1$. The choice of $t_1(\bullet)$ and $t_2(\bullet)$ will affect the power but not the level of the test. This procedure can be implemented in a randomized trial where $f(A_1 \mid L, A_0)$ is known by design. In observational studies, in a preliminary step, one must specify a parametric model $f(a_1 \mid \ell, a_0; \alpha)$ and find $\widehat{\alpha}$ that maximizes the likelihood $\prod_{i=1}^{n} f(A_{1i} \mid L_i, A_{0i}; \alpha)$ and replace $\mathcal{W}$ by $\mathcal{W}(\widehat{\alpha}) = t_1(Y) t_2(A_1) / f(A_1 \mid L, A_0; \widehat{\alpha})$. When $\alpha$ is estimated, the $p$-value outputted by off-the-shelf software will exceed the true $p$-value (i.e., the test is conservative), although a corrected $p$-value can be easily computed (Robins, 1997b).

### 8.3 Direct-effect SNDMs

We now generalize this example by considering estimation and testing of direct effects using direct-effect SNDMs. We suppose that treatment $A_m = (A_{Pm}, A_{Zm})$ at time $t_m$ is comprised of two different treatments $A_{Pm}$ and $A_{Zm}$. To formalize the no-direct-effect null hypothesis, let $g_p \equiv (g_{P0}, \ldots, g_{PK})$ be a collection of functions where $g_{Pm} : \overline{\mathcal{L}}_m \to \mathcal{A}_{Pm}$. Then, for history $\overline{a}_Z \equiv \overline{a}_{ZK} \in \overline{\mathcal{A}}_Z$, let $g = (g_P, \overline{a}_Z)$ be the treatment regime or plan given by $g_m(\bar{\ell}_m) = \{g_{Pm}(\bar{\ell}_m), a_{Zm}\}$. Then $F_{(g_P, \bar{a}_Z)}(y)$ is the distribution of $Y$ that would be observed if $\bar{A}_Z$ was set to $\bar{a}_Z$ and the treatments $\bar{A}_P$ were assigned, possibly dynamically, according to the plan $g_P$. If $g_P$ is the non-dynamic regime $(\bar{a}_{Pm}, 0)$, we write $F_{(\bar{a}_{Pm}, \bar{a}_Z)}(y)$.

**Definition:** The direct effect "g"-null hypothesis of no direct effect of $\bar{A}_P$ controlling for $\bar{A}_Z$ is $F_{g_P, \bar{a}_Z}(y) = F_{g_{P*}, \bar{a}_Z}(y)$ for all $\bar{a}_Z, g_P, g_{P*}$.

Let $\gamma(y, \bar{\ell}_m, \bar{a}_{Pm}, \bar{a}_Z)$ be the quantile-quantile function mapping quantiles of $F_{(\bar{a}_{Pm}, \bar{a}_Z)}(y \mid \bar{\ell}_m)$ into quantiles of $F_{(\bar{a}_{P(m-1)}, \bar{a}_Z)}(y \mid \bar{\ell}_m)$ which satisfies (a) $\gamma(y, \bar{\ell}_m, \bar{a}_{Pm}, \bar{a}_Z) = y$ if $a_{Pm} = 0$; (b) $\gamma(y, \bar{\ell}_m, \bar{a}_{Pm}, \bar{a}_Z)$ is increasing in $y$; and (c) the derivative of $\gamma(y, \bar{\ell}_m, \bar{a}_{Pm}, \bar{a}_Z)$ w.r.t. $y$ is continuous.

Robins (1997b) proved that $\gamma(y, \bar{\ell}_m, \bar{a}_{Pm}, \bar{a}_Z) \equiv y$ if and only if the direct-effect "g"-null hypothesis holds. We now construct a parametric model for $\gamma(y, \bar{\ell}_m, \bar{a}_{Pm}, \bar{a}_Z)$.

**Definition:** The distribution $F$ of $V$ follows a direct-effect pseudo-structural nested distribution model $\gamma(y, \bar{\ell}_m, \bar{a}_{Pm}, \bar{a}_Z, \psi)$ if $\gamma(y, \bar{\ell}_m, \bar{a}_{Pm}, \bar{a}_Z) = \gamma(y, \bar{\ell}_m, \bar{a}_{Pm}, \bar{a}_Z, \psi_0)$ where (1) $\gamma(\cdot, \cdot, \cdot, \cdot, \cdot)$ is a known function; (2) $\psi_0$ is a finite vector of unknown parameters; (3) for each value of $\psi$, $\gamma(y, \bar{\ell}_m, \bar{a}_{Pm}, \bar{a}_Z, \psi)$ satisfies the above conditions (a)-(c); (4) $\partial \gamma(y, \bar{\ell}_m, \bar{a}_{Pm}, \bar{a}_Z, \psi)/\partial \psi'$ and $\partial^2 \gamma(y, \bar{\ell}_m, \bar{a}_{Pm}, \bar{a}_Z, \psi)/\partial \psi' \partial y$ are continuous; and (5) $\gamma(y, \bar{\ell}_m, \bar{a}_{Pm}, \bar{a}_Z, \psi) = y$ if and only if $\psi = 0$ so that $\psi_0 = 0$ represents the direct-effect "g-" null hypothesis.

We now consider testing and estimation of $\psi_0$. Our fundamental tool is the following theorem of Robins (1997b) characterizing $\gamma(y, \bar{\ell}_m, \bar{a}_{Pm}, \bar{a}_Z)$. For any function $\gamma^*(y, \bar{\ell}_m, \bar{a}_{Pm}, \bar{a}_Z)$ satisfying conditions (a)-(c) above, we recursively redefine the following: $H_K(\gamma^*) = \gamma^*(Y, \bar{L}_K, \bar{A}_{PK}, \bar{A}_Z)$, $H_m(\gamma^*) = \gamma^*(H_{m+1}(\gamma^*), \bar{L}_m, \bar{A}_{Pm}, \bar{A}_Z)$, and set $H(\gamma^*) \equiv H_0(\gamma^*)$. Define $W_m = \prod_{k=m}^{K} f(A_{Zk} \mid \bar{A}_{k-1}, \bar{L}_{k-1})$ and $\underline{A}_{Zm} = (A_{Zm}, \ldots, A_{ZK})$.

**Theorem 5:** $\gamma^*(Y, \bar{L}_m, \bar{A}_{Pm}, \bar{A}_Z) = \gamma(Y, \bar{L}_m, \bar{A}_{Pm}, \bar{A}_Z)$ w.p.1 if and only if for $m = 0, \ldots K$ and any functions $t_m(\cdot, \cdot)$, $E\left[t_m\left(\underline{A}_{Z(m+1)}, H(\gamma^*)\right)/W_{m+1} \mid \bar{A}_m, \bar{L}_m\right]$ does not depend on $A_{Pm}$ w.p.1, when the expectation is finite.

Given a direct-effect SNDM, define $H(\psi)$ to be $H(\gamma^*)$ with $\gamma^*$ the function $\gamma(y, \bar{\ell}_m, \bar{a}_{Pm}, \bar{a}_Z, \psi)$. Theorem 5 implies that we can construct tests and confidence intervals for $\psi_0$ in observational studies using off-the-shelf software as follows.

Step 1: Specify a parametric model $f\left(A_{Zk} \mid \bar{A}_{k-1}, \bar{L}_k; \alpha^{(1)}\right)$ and calculate the MLE $\hat{\alpha}^{(1)}$ that maximizes $\prod_i \prod_{k=0}^{K} f\left(A_{Zki} \mid \bar{A}_{(k-1)i}, \bar{L}_{ki}; \alpha^{(1)}\right)$ and let $W_m\left(\hat{\alpha}^{(1)}\right)$ denote $W_m$ evaluated under the density indexed by $\hat{\alpha}^{(1)}$.

Step 2: For $m = 0, \ldots, K$, specify a model for the conditional mean of $A_{Pm}$ depending on $\alpha^{(0)}$

$$E\left[A_{Pm} \mid A_{Zm}, \bar{A}_{m-1}, \bar{L}_m\right] = d\left(\alpha^{(0)'} Q_m\right) \quad (28)$$

where $Q_m$ is a known vector function of $\bar{A}_{Zm}, \bar{A}_{m-1}, \bar{L}_m$ and $d(\bullet)$ is a known link function. For example, if $A_{Pm}$ is dichotomous, we might choose $d(x) = \{1 + \exp(-x)\}^{-1}$.

Step 3: Compute an $\alpha$-level test of the hypothesis that $\theta = 0$ in the extended model that adds the term $\theta' Q_m^*(\psi) = \theta' q_m^*\left(H(\psi), \bar{A}_{m-1}, \bar{L}_m, \underline{A}_{Zm}\right)/W_{m+1}\left(\hat{\alpha}^{(1)}\right)$ to the $\alpha^{(0)'} Q_m$, where (i) $q_m^*(\bullet)$ is a chosen function of the dimension of $\psi$, (ii) in testing $\theta = 0$, we treat the $Q_m^*(\psi)$ as "fixed covariates" and (iii) we use generalized estimating equation (GEE) software available in S+ or SAS that regards $A_{P0}, \ldots, A_{PK}$ as correlated.

This test is a conservative $\alpha$-level test of the hypothesis $\psi = \psi_0$. A conservative 95% confidence interval, guaranteed to cover $\psi_0$ at least 95% of the time in large samples, is the set of $\psi$ for which the .05-level test of $\theta = 0$ fails to reject. The tests and interval are conservative because standard software programs do not adjust for the effect of estimating $\alpha^{(1)}$.

Robins (1997b) describes a complete reparameterization of the distribution of $V$ with the direct-effect SNDM model $\gamma(y, \bar{\ell}_m, \bar{a}_{Pm}, \bar{a}_Z, \psi)$ as a component and describes how to estimate, with this reparameterization, the contrasts $F_{(g_P, \bar{a}_Z)}(y) - F_{(g_{P*}, \bar{a}_Z)}(y)$.

**Appendix 1: Proof of (18):**
We will show by induction that

$$pr\left[H_m > y \mid \bar{L}_m, \bar{A}_m\right] = F_{g=(\bar{A}_{m-1}, 0)}\left(y \mid \bar{L}_m\right) \quad (29)$$

which implies $A_m \prod H_m \mid \bar{L}_m, \bar{A}_{m-1}$. Furthermore, $H$ is a deterministic function of $(H_m, \bar{L}_m, \bar{A}_{m-1})$ which proves (18).

**Case 1:** $m = K$: $pr\left[H_K > y \mid \bar{L}_K, \bar{A}_K\right] = pr\left[Y > \gamma^{-1}(y, \bar{L}_K, \bar{A}_K) \mid \bar{L}_K, \bar{A}_K\right] \equiv F_{g=(\bar{A}_K)}\left[\gamma^{-1}(y, \bar{L}_K, \bar{A}_K) \mid \bar{L}_K\right] \equiv F_{g=(\bar{A}_{K-1}, 0)}\left(y \mid \bar{L}_K\right)$.

**Case 2:** Assume (A.1) holds for $m$. If we can show it holds for $m - 1$, Eq. (18) is proved. $pr[H_{m-1} > y \mid \bar{L}_{m-1}, \bar{A}_{m-1}] = \int \{pr[H_m > \gamma^{-1}(y, \bar{L}_{m-1}, \bar{A}_{m-1}) \mid \bar{L}_m, \bar{A}_m]\} dF[L_m, A_m \mid \bar{L}_{m-1}, \bar{A}_{m-1}] = \int F_{g=(\bar{A}_{m-1}, 0)}[\gamma^{-1}(y, \bar{L}_{m-1}, \bar{A}_{m-1}) \mid \bar{L}_m] dF(L_m, A_m \mid \bar{L}_{m-1}, \bar{A}_{m-1}) = F_{g=(\bar{A}_{m-1}, 0)}[\gamma^{-1}(y, \bar{L}_{m-1}, \bar{A}_{m-1}) \mid \bar{L}_{m-1}] \equiv F_{g=(\bar{A}_{m-2}, 0)}(y \mid \bar{L}_{m-1})$ where the third to last equality is by the induction hypothesis and the second to last by the definition of $F_{g=\overline{a}}(y \mid \overline{L}_k)$.